\documentstyle[PASJadd,epsf]{PASJ95}

\newcommand{\vol}{51}
\newcommand{\no}{4}
\newcommand{\lett}{}
\newcommand{\spage}{000}
\newcommand{\rdate}{1998 November 02}
\newcommand{\adate}{1999 May 17}

\begin{document}
\setcounter{page}{000}

\title{Metal Enrichment in the Early Galactic Halo}

\author{Chisato {\sc Ikuta} and Nobuo {\sc Arimoto}\\
{\it Institute of Astronomy, School of Science, University of Tokyo,
Mitaka, Tokyo 181-8588, JAPAN } \\
{\it E-mail(CI): ikuta@mtk.ioa.s.u-tokyo.ac.jp}
}

\abst{
An early history of metal enrichment in the Galactic halo is studied. 
We investigate chemical inhomogeneity by using a stochastic chemical 
evolution model. The model confronts with metallicity distribution function 
of long-lived halo stars which is found to be a clue to obtain the best 
model prescriptions. 
We find that the star formation in the halo virtually terminated by 
$\sim 1$ Gyr and that the halo has never been chemically homogeneous 
in its star formation history. 
This conclusion does not depend whether mass loss from the halo is taken 
into account or not. Observed ratios of $\alpha$-elements with 
respect to iron do not show scatters on a [$\alpha$/Fe]-[Fe/H] 
plane, but this does not imply that interstellar matter in the halo 
was homogeneous because a chemical evolution path on this diagram 
is degenerate in the star formation rate. On the other hand, 
apparent spread of [Sr/Fe] ratio among 
metal-poor halo stars does not reflect an inhomogeneous metal enrichment, 
instead it is due to a sharp increase in a production rate of 
strontium that is probably synthesised in slightly less massive stars 
than progenitor of iron-producing SN\,II. }

\maketitle
\input epsf



\thispagestyle{headings}

\section{Introduction \label{sec:intro}}
At very beginning of enrichment history of the Galactic halo,
interstellar medium (ISM) was certainly chemically inhomogeneous, 
because star forming regions at that time were locally confined 
and stellar ejecta did not spread uniformly throughout the halo.
Only a limited number of Type II supernovae (SN\,II) exploded
and a mean spatial separation of SN\,II was much larger than 
a typical radius of single supernova remnant (Audouze \& 
Silk 1995). Thus, the ISM was enriched locally. 
A trace of such inhomogeneous ISM should be frozen 
in extremely metal-deficient 
stars which we observe today (e.g. McWilliam 1997).

Nevertheless, no clear evidence which suggests  
that the halo was inhomogeneous in the past has so far been found. 
McWilliam et al. (1995), Audouze \& Silk (1995), 
and McWilliam (1997) claimed that large scatters of {\it heavy metal}  
abundances (e.g., Sr, Y, Zr, Ba, and Eu) 
of extremely metal-poor halo stars (Beers, Preston, \& Shectman 1992; 
McWilliam et al. 1995; 
Ryan, Norris, \& Beers 1996) imply that the early halo was inhomogeneous. 
However, these scatters may 
simply reflect various lifetime of stars producing heavy metals, 
whose origins are poorly understood yet. 
The abundance of each heavy metal should sharply rise at a time when 
bulk of massive stars that produce it start to explode. Thus stars formed 
during such epoch should show a large scatter in relative abundance of 
heavy metals with 
respect to iron while they have nearly identical iron content. 
Therefore large scatters of heavy metals at certain values of [Fe/H] do 
not necessarily imply an inhomogeneous ISM in an early halo. 

On the other hand, even if relative abundance of $\alpha$-elements 
(e.g., O, Mg, Si, Ca, etc) with respect to iron do show little scatter 
at any value of [Fe/H], it does not necessarily mean that an early 
halo was homogeneous. Even if a star formation history of each region 
in the halo is different from each other, evolutionary paths on the 
[$\alpha$/Fe]-[Fe/H] plane should be nearly identical in an early 
evolutionary stage during which SN\,II were dominant sources of nuclear 
enrichment. The evolutionary path depends on a star formation rate (SFR) 
very weak. 

Although little is known about an early halo, there seems to be a clear 
evidence for the inhomogeneous ISM in a solar neighbourhood disc; 
long-lived dwarfs (e.g., Edvardsson et al 1993) show a spread of $\sim1$\,dex 
in iron abundance along poorly defined age-metallicity relation. 
By relaxing a usual assumption of well mixing, Copi (1997) estimated 
chemical evolution of the solar neighbourhood in a Monte Carlo fashion. 
The solar neighbourhood is modeled by 1000 independent 
regions. The evolutionary history of one particular region is determined 
randomly based on the SFR and the initial mass function (IMF). 
Although he successfully reproduced the observed spreads of  elemental ratios, 
he failed to explain the scatter appeared in the age-metallicity relation. 
Bateman \& Larson (1993) suggested that random walk processes of atomic 
and molecular gas clouds are dominant mixing mechanisms of iron in the 
present day solar neighbourhood disc. Wielen, Fuchs, \& Dettbarn (1996) 
suggested that stellar orbital diffusion in combination with radial abundance 
gradients can induce inhomogeneity in the ISM. However, van den Hoek \& 
de Jong (1997) showed that the stellar orbital diffusion  
does not explain the abundance variation sufficiently. They suggested instead 
that a sequential star formation and an infall of metal-deficient gas 
play an important role in preventing the ISM from being mixed. 
Wilmes \& K\"oppen (1995) studied chemical evolution of isolated 
individual ISM parcels and showed that mixing is inefficient 
in the galactic disc. Pilyugin (1996) suggested that major galaxy mergers
form multiple stellar populations of different metallicities in the disc. 
While these mechanisms can well explain the scatters along the age-metallicity 
relation, one should also keep in mind that stellar abundances may not reflect 
abundances of ISM from which stars formed, 
because chemical condensation processes 
such as grain formation in circumstellar envelope (e.g. Henning \& 
G\"urler 1986) and/or thermal diffusion in stellar 
atmosphere (e.g. Bahcall \& Pinsonneault 1996) might work efficiently.

If the early Galactic halo was chemically inhomogeneous, the metallicity 
spread should appear among halo stars of the same age. Unfortunately, 
the present day stellar isochrone fitting cannot distinguish age of 
stars born in a young halo. Therefore, it is not clear at all if the 
metallicity spread exists among coeval halo stars. Instead, we show 
in this paper that one can use an observed cumulative metallicity 
distribution function of halo stars to constrain stochastic chemical 
evolution models for an early halo and indicate that the Galactic halo 
has never been chemically homogeneous in its history. 

In section 2, we give prescriptions for our stochastic chemical evolution 
models, and in section 3 we describe behaviours of theoretical metallicity 
distributions. We confront model results with observational data in section 4 
and give discussions and conclusions in section 5 and 6, respectively. 

\section{Model \label{model}}
\subsection{Outline} 
We have built up stochastic chemical evolution models for the Galactic halo. 
We assume that the halo is spheroidal and that the gas of primordial 
abundance distributed uniformly at the beginning. We divide the halo into 
many cubic blocks having a volume of $l_{\rm b}^3=4 \pi R_{\rm m}^3/3$, 
where $R_{\rm m}$ 
is a radius of supershell caused by SN\,II that contributed dominantly 
to an early stage of chemical evolution. Progenitor of SN\,II are massive 
stars which tend to form in star clusters and associations 
(Blaauw 1964; Humphreys 1978; Heiles 1987).      
Supernova remnants in OB associations and star clusters are suggested 
to produce surrounding shell structures, or supershells 
(e.g., Tenorio-Tangle \& Bodenheimer 1988). 
Cash et al. (1980) predicted that the supershell is produced after a series 
of SN\,II explosions.  A radius of supershell can roughly be given as 
a maximum size of the area enriched by synthesised heavy elements. 
Thus we assume that a volume of the block is equal to that of the supershell. 
The ISM in the supershell should be well-mixed, because 
stellar winds and SN\,II explosions induce 
the Rayleigh-Taylor and the Kelvin-Helmholtz instabilities
(Tenorio-Tagle \& Bodenheimer 1988; Allen \& Burton 1993). 
The time scale of mixing is less than 0.0015 Gyr in a case of isonised gas 
(Roy \& Kunth 1995), thus is much shorter than a typical lifetime 
($\sim 0.01$\,Gyr) of the OB association. Once the mixing took place, 
the ISM in a block uniformly enriched by newly synthesised elements. 
Therefore, individual regions can be regarded as independent one-zones and 
a standard chemical evolution model can be applied for each of them. 

Before an onset of star formation, all blocks have identical physical 
properties. We therefore follow chemical enrichment histories of 
1000 neighbouring blocks instead of the all. Unless a number of blocks 
considered is too small, the resulting behaviour of each block is 
sufficiently stable. By simulating stochastic chemical evolution for 1000 
blocks simultaneously, we calculate time variations of abundances in 
individual blocks and study chemical inhomogeneity in the halo. 

\subsection{Block Size}
In an analogy to an expansion of single supernova, 
we roughly estimate a radius of the supershell as follows, 
although more detailed models for the supershells were 
published by several authors 
(Bruhweiler et al. 1980; McCray \& Kafatos 1987; Mac Low \& McCray 1988). 
Cioffi, McKee, \& Bertschinger (1988) gave an analytical 
estimate for a radius $r_{\rm m}$ of a supernova remnant (SNR): 
\begin{equation}
r_{\rm m}=69~E_{51}^{0.32}~n^{-0.41}~v_{10}^{-0.43}~(Z/Z_\odot)^{-0.05}~~{\rm{pc}}, 
	\label{eqn:Rm}
\end{equation}
where $E_{51}$ is an initial kinetic energy of the SNR 
in units of $10^{51}$ erg, $n$ is the number density of surrounding ISM, 
$v_{10}$ is an ISM velocity dispersion in units of 10 km s$^{-1}$, 
and $Z/Z_\odot$ is the ISM heavy element abundance. We calculate a radius 
$R_{\rm m}$ of the supershell by
replacing $E_{51}$ with $n_{\rm{II}}\,E_{51}$, where $n_{\rm{II}}$ is the number of SN\,II in an OB association. 
Hereafter we adopt $E_{51}=1$ (Shigeyama, Nomoto, \& Hashimoto 1988). 
We assume that the velocity dispersion of ISM is equal to the sound velocity,  
and adopt $v \simeq 10$~km\,s$^{-1}$ or $v_{10} \simeq 1$, since  
the temperature should be near $10^4$~K (Hoyle 1953; Silk 1977) in the 
initial halo. 
Since we are interested in chemical evolution in the halo, 
we fix $Z/Z_\odot=0.06$ which corresponds to 
[O/Fe]$\simeq 0.4$ (Barbuy 1988; Nissen et al. 1994) and    
[Fe/H]$=-1.6$; i.e., a peak iron abundance 
of the metallicity distribution function 
obtained for long-lived stars in the halo (Laird et al. 1988). 

Stars having $M> 8 M_\odot$, corresponding to main-sequence spectral type 
B3, eventually become SN\,II. The number of supernovae $n_{\rm{II}}$ 
is the number of such stars per OB association. We adopt $n_{\rm{II}}=40$ 
according to Heiles (1987) who derived the average number of stars 
with $M> 8 M_\odot$ per clusters from actual counting of O stars in clusters 
in the solar neighbourhood together with the IMF derived from clusters in the 
solar vicinity. The number Heiles (1987) got was $\sim 28$, which was then 
corrected for runaway O stars that will deposit energy within the supershell. 
Significant deviations from the average value of $n_{\rm{II}}$ are 
apparently rare although $n_{\rm{II}}$ may cover a wide range of values 
(Tenorio-Tangle \& Bodenheimer 1988). 

The number density of the ISM is given as 
$n=M_{\rm h}/(\mu m_{\rm H} V_{\rm h})$, where $m_{\rm H}$ is the proton mass, 
 $M_{\rm h}$ is the mass, $V_{\rm h}=\frac{4}{3} \pi R_{\rm h}^3$ is 
the halo volume, and $\mu=1.3$ is a mean molecular weight corresponding 
to the primordial compositions.

Saito (1979) derived an empirical relation between   
the binding energy $\Omega_{\rm G}$ and the mass $M_{\rm G}$ 
by analysing surface brightness distributions
and line-of-sight velocity dispersions of spheroidal systems:   
\begin{equation}
\Omega_{\rm G}=1.66 \times 10^{60} \left[\frac{M_{\rm G}}
{10^{12}M_\odot} \right]^{1.45}~~~{\rm{erg}}.  
\end{equation}
Under the assumption of spherical geometry, 
the radius $R(M_{\rm G})$ of a galaxy is given as 
\begin{equation}   
R(M_{\rm G})
=26.1 \left[\frac{M_{\rm G}}{10^{12}~M_\odot} \right]^{0.55}~~\rm{kpc}.
	\label{eqn:Rh}
\end{equation}
We assume that the early Galactic halo follows 
the same mass-radius relation given by equation (\ref{eqn:Rh}). 
The virial theorem tells that a system with initially no kinetic energy 
attains virial equilibrium by reducing a radius 
to half the initial value.  Since we study chemical evolution 
at the very beginning, the radius of halo $R_{\rm h}$ 
should be taken as $2R(M_{\rm h})$. 

If we assume the halo was $M_{\rm h}= 4 \cdot 10^{11}~M_\odot$ 
(Fich \& Tremaine 1991), then we obtain $R_{\rm h}=2R(M_{\rm h})=31.5$~kpc, 
and $n=0.1$cm$^{-3}$. According to equation (1), the supershell radius 
$R_{\rm m}$ depends weakly on $n$. Since the amount of gas condensed into 
stars in the early stage of halo evolution is at most 25 percent, 
$R_{\rm m}$ changes only by 12 percent. We therefore assume $n$ constant 
in time. Putting $n_{\rm{II}}E_{51}=40$, $n=0.1$cm$^{-3}$, 
$v_{10} \simeq 1$, and $Z/Z_\odot=0.06$ into equation (1), we obtain 
$R_{\rm m} \sim 660$~pc and $l_{\rm b} \equiv (4\pi R_m^3/3)^{1/3}=1~$kpc 
for our standard model. 

A time $t_{\rm m}$ when a SNR merges with the surrounding ISM
is given by (Cioffi et al. 1988), 
\begin{equation}
t_{\rm m}=2 \cdot 10^6E_{51}^{0.32}~n^{-0.37}~v_{10}^{-1.43}~(Z/Z_\odot)^{-0.05}~~{\rm{yr}},
	\label{eqn:Tm}
\end{equation}
and a cooling time  $t_{\rm c}$ inside a SNR (Cox 1972) as, 
\begin{equation}
t_{\rm c}=5.7 \cdot 10^4~E_{51}^{4/17}~n^{-9/17}~~{\rm{yr}}.  
	\label{eqn:tc}
\end{equation}
A lifetime of supershell $T_{\rm m}$ and a cooling time $T_{\rm c}$ 
inside the supershell can be derived by replacing $E_{51}$ in equations (4) 
and (5) with $n_{\rm II}E_{51}$, respectively. Assuming the same parameters 
discussed above, we obtain $T_{\rm m} \simeq 1.7 \cdot 10^{-2}$~Gyr and 
$T_{\rm c} \simeq 4.6 \cdot 10^{-4}$\,Gyr. We note that $T_{\rm m}$ 
is much longer than $T_{\rm c}$. 

\subsection{Star Formation Probability}
Considering an idealised situation that the OB associations distribute 
uniformly in the halo, we assume that a star formation probability 
$P_{\rm{SF}}(t)$ in a block for a time interval $t$ and $t+\Delta t$ 
is given as,  
\begin{eqnarray}
P_{\rm{SF}}(t) & = & \frac{l_{\rm b}^3}{V_{\rm h}} \frac{1}{n_{\rm{II}}}
\int^{t+\Delta t}_{t}N_{\rm{II}}(t') dt' \nonumber \\
	& & \nonumber \\
       & = & \frac{3}{4 \pi} \left(\frac{l_{\rm b}}{R_{\rm h}}
 \right)^3 \frac{1}{n_{\rm{II}}} \int^{t+\Delta t}_t N_{\rm{II}}(t') dt', 
\end{eqnarray}
where $N_{\rm{II}}(t)$ is the total number of progenitor 
of SN\,II in the halo and is given as,  
\begin{equation}
N_{\rm{II}}(t)
=M_{\rm h} C(t) f_{\rm{II}},  
	\label{eqn:nsfb}
\end{equation}
where $C(t)$ is the SFR per unit mass and $f_{II}$ is the number fraction 
of SN\,II defined as, 
\begin{equation}
f_{\rm{II}}= \int_{m_{\rm{II}}}^{m_u} \phi(m) m^{-1}dm,
\end{equation}
with $m_l=0.1~M_\odot$, $m_u=50~M_\odot$, and $m_{\rm{II}}=10~M_\odot$  
as the lower and the upper mass limits, 
and the lower SN\,II mass limit, respectively. The IMF adopted here has 
mass spectrum independent of time:
\begin{equation}
\phi(m)=\frac{(x-1)~m_l^{x-1}}{1-(m_l/m_u)^{x-1}}~m^{-x},~~(m_l\le m \le m_u). 
	\label{eqn:imf}
\end{equation}
We adopt the Salpeter IMF (Salpeter 1955) which has a slope $x=1.35$ 
in this definition. For $C(t)$, we adopt the Schmidt law (Schmidt 1959):   
\begin{equation}
C(t)=\omega^{-1}~G(t)^{\,p}~~~~~(p=1),
\end{equation}
where $\omega$ and $G(t)$ are time scale of star formation 
and a gas fraction in the halo, respectively.  
 $P_{\rm{SF}}(t)$ is thus finally given as:
\begin{eqnarray}
\lefteqn{P_{\rm{SF}}(t) = \frac{3}{4 \pi}\,\left(\frac{l_{\rm b}}{R_{\rm h}}
\right)^3 n_{\rm{II}}^{-1} f_{\rm{II}}
\,M_{\rm h}\, \omega^{-1} G(t)^p\, \Delta t}\nonumber \\
&& \hspace{4cm}(p=1). 
	\label{eqn:psf} 
\end{eqnarray}

We adopt $m_{\rm II}=10M_\odot$ in equation (8), which is slightly larger than 
$m_{\rm II}=8M_\odot$ which Heiles (1987) adopted.  
Thus, equation (11) may underestimate $P_{\rm{SF}}(t)$. 
A difference corresponding to the adopted $m_{\rm II}$ is 
$\sim 12$ percent. 
However, an exact value of $m_{\rm II}$ is still uncertain and, therefore, 
we adopt $m_{\rm II}=10M_\odot$ for our standard model and change it later 
as a free parameter. 

\subsection{Chemical Evolution}
For the blocks in which stars are born (hereafter the star forming blocks),  
we apply a chemical evolution model developed by 
Pagel \& Tautvai$\check{{\rm s}}$ien$\dot{{\rm e}}$ 
(1995; hereafter PT95). We adopt the iron as a tracer of chemical evolution. 
In PT95, the iron abundance is calculated 
by using an instantaneous recycling approximation 
and a delayed production approximation  
which assumes that the iron is additionally produced  
by Type Ia supernovae (SNIa) after a fixed time lag $\tau_1$.
We assume that the SFR $c(t)$ in a block is proportional 
to the gas fraction $g(t)$. The iron abundance $Z_{\rm{Fe}}(t)$ is given by, 
\begin{footnotesize}
\begin{eqnarray} 
	\label{eqn:delay}
\lefteqn{\frac{d(gZ_{\rm{Fe}})}{dt}}\nonumber  \\
&&= \left\{
   \begin{array}{ll}
	-Z_{\rm{Fe}}(t)c(t)+p_0c(t), & (t < \tau_1),\\  
        -Z_{\rm{Fe}}(t)c(t)+p_0c(t)+p_1c(t-\tau_1), & (t \ge \tau_1),
    \end{array}
 \right. 
\end{eqnarray}
\end{footnotesize}
where $p_0$ and $p_1$ are yields of iron corresponding to the instantaneous 
recycling approximation and the delayed production approximation, 
respectively.  
The first term on the right hand side (r.h.s.) represents 
the net amount of iron locked into newly formed stars. 
The second and the third terms are 
corresponding to the instantaneous recycling approximation
and the delayed production approximation, respectively. 
We adopt $p_0=0.28$, $p_1=0.42$, and $\tau_1=1.3$~Gyr   
according to PT95. 
Since the theoretical yields of the iron are uncertain due to difficulties 
in modeling the explosion mechanism (Timmes, Woosley, \& Weaver 1995), 
we use the yields of PT95 which are calibrated empirically. 
  
The SFR in each block $c(t)$ is determined by an assumed number of SN\,II 
exploded in a block during a time interval $t$ and $t + \Delta t$:   
\begin{equation}
\int^{t+ \Delta t}_{t}m_{\rm b} f_{\rm{II}} c(t')~dt'   
=n_{\rm{II}},   
\end{equation}
where $m_{\rm b}= M_{\rm h} l_{\rm b}^3/V_{\rm h} 
\simeq 3 \cdot 10^6\,M_\odot$ is the initial gas mass of each block. 

\subsection{Standard Models}
Table\,1 gives a list of parameters adopted for our standard models\,S1, 
S2, and S3. In model\,S1 we assume that the OB associations are formed 
randomly at every $\Delta t \sim \frac{1}{3}T_{\rm m}=0.006$\,Gyr. 
In model S2 we introduce an effect of periodic mixing due to the 
turbulent diffusion. A particle with a sound velocity $v=10 {\rm km\,s^{-1}}$ 
would across a block of size $l_{\rm b}=1$\,kpc in $T_{\rm d}=0.1$\,Gyr. 
Thus, we assume in this model that every 0.1\,Gyr the turbulent diffusion 
mixes the 27 blocks surrounding the star forming block periodically. 

\begin{table}[t]
\small
\begin{center}
Table~1. \hspace{4pt}Parameters of standard model.\\
\end{center}
\vspace{6pt}
\begin{tabular*}{\columnwidth}{@{\hspace{\tabcolsep}
\extracolsep{\fill}}lll}
\hline\hline\\[-6pt]
$M_{\rm h}$ & $4~10^{11}~M_\odot$ & mass of the halo\\
$R_{\rm h}$ & 31.5~kpc & radius of the halo\\
$\omega$ & 5~Gyr & time scale of star formation in the halo \\
$l_{\rm b}$ & 1~kpc & size of a block\\
$v_{10}$ & 1 & velocity dispersion of ISM \\
 & & \hspace{2cm} in units of 10 km~s$^{-1}$ \\
$m_l$ & 0.1$M_\odot$ & IMF lower mass limit \\
$m${\tiny II} & 10$M_\odot$ & lower mass limit for SN\,II\\
$n_{\rm{II}}$ & 40 & number of SNe\,II in a block\\
$\Delta t$ & $6 \cdot 10^6$~yr & duration of star formation\\
\hline
\end{tabular*}
\end{table}

 The supershell structures are often mentioned as a trigger of massive star 
formation. The evidences have been discussed by many authors  
(Blaauw 1964; Elmegreen \& Lada 1977; 
Lada, Blitz, and Elmegreen 1979; Elmegreen 1982; 1985a, b).
If OB associations form in this way, 
they should be born only in the surrounding area of the 
star forming regions and chemical enrichment should be locally confined 
there. The effects of this stimulated star formation are studied in model~S3. 
We first distribute OB associations randomly and then assign randomly 
a star forming block for the next generation among 27 blocks adjoining to 
each star forming block of the first generation and so on. In this way, 
the star formation propagates from one block to another. 

The time step adopted for each simulation is 0.001\,Gyr. Calculations are 
stopped at 1\,Gyr, since our main interest is in early chemical evolution 
of the halo.  

The time scale of star formation $\omega$ in the halo 
is quite uncertain. On the contrary, it is quite often assumed that
$\omega$ in elliptical galaxies is of the order
of free-fall time $t_{\rm ff}=(3 \pi / 32G \bar{\rho})^{1/2}$
(Larson 1974). One could assume that $\omega$ in the halo is the same
as that in giant ellipticals. However, mean stellar metallicities
of giant ellipticals are nearly solar (Arimoto et al. 1997), 
while that of Galactic halo stars  
is $\sim$ 1/40 solar (Ryan \& Norris 1991; Carney et al. 1996).
Both in giant ellipticals and in the galactic halo, Mg seems to
be enhanced with respect to Fe by at least a factor of 2
(McWilliam et al 1995; Ryan et al 1996). 
This implies that star formation in both systems is  
characterised by the same IMF and that the star formation stopped
before the onset of SNIa explosions ($\sim 1$ Gyr).
If this is the case, $\omega$ in the Galactic halo 
should be 40 times longer than that of giant ellipticals.
Thus we obtain $\omega \simeq 40 t_{\rm ff} \sim 6$\,Gyr 
for our halo model with $M_{\rm h}= 4 \cdot 10^{11}~M_\odot$ and 
$R_{\rm h}=31.5$~kpc.
Since $\omega \sim 6$\,Gyr is very close to $\omega  
\sim 5$\,Gyr derived for the solar neighbourhood disc (Arimoto, 
Yoshii, \& Takahara 1992), we shall adopt $\omega=5$\,Gyr in our standard 
models. In other words, we adopt the same SFR fo the Galactic
halo and the solar neighbourhood disc.
  
\section{Metallicity Distribution \label{subsubsec:sk}}

Figure~1 shows the number fraction of blocks 
which at least once have experienced the chemical enrichment. Model\,S1 
predicts  that all the blocks have been enriched before $\sim 0.2$~Gyr. 
In model\,S2, this happens earlier than model\,S1,  because the iron 
spreads beyond the supershells due to the turbulent diffusion, 
thus much wider area are enriched even if the SFRs 
are the same as in model\,S1. On the other hand, 
nearly 8\% 
of the block in model\,S3 has never been enriched till 1~Gyr, 
this is because the stimulated star formation 
tends to localise the chemical enrichment. 
\begin{figure}
\epsfxsize=150pt
\epsfbox{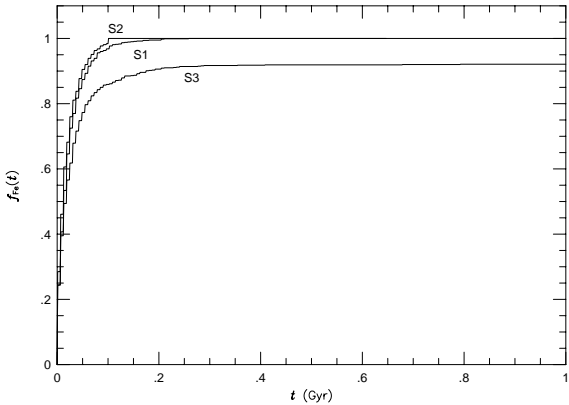}
\caption[]{The evolution of the number fraction of blocks 
which at least once have experienced chemical enrichment 
in the standard models\,S1, S2, and S3. 
}
\end{figure}

Figures\,2a~and~2b show the evolution of mean and median iron abundances, 
respectively. 
Hereafter we discuss statistical properties 
except for the first $0.04$\,Gyr, during which the number of 
enriched blocks is too small to define the statistical properties.   
Figures\,2a~and~2b indicate that the median metallicities of 
models\,S1~and~S2 are nearly the same as the mean metallicities. 
This suggests that the metallicity distributions of the two models 
have symmetric shapes. The median of model\,S3 is always higher 
than the mean, indicating that the metallicity distribution has a tail 
toward lower metallicity.

\begin{figure}[h]
\begin{center}
\epsfxsize=9cm
\epsfbox{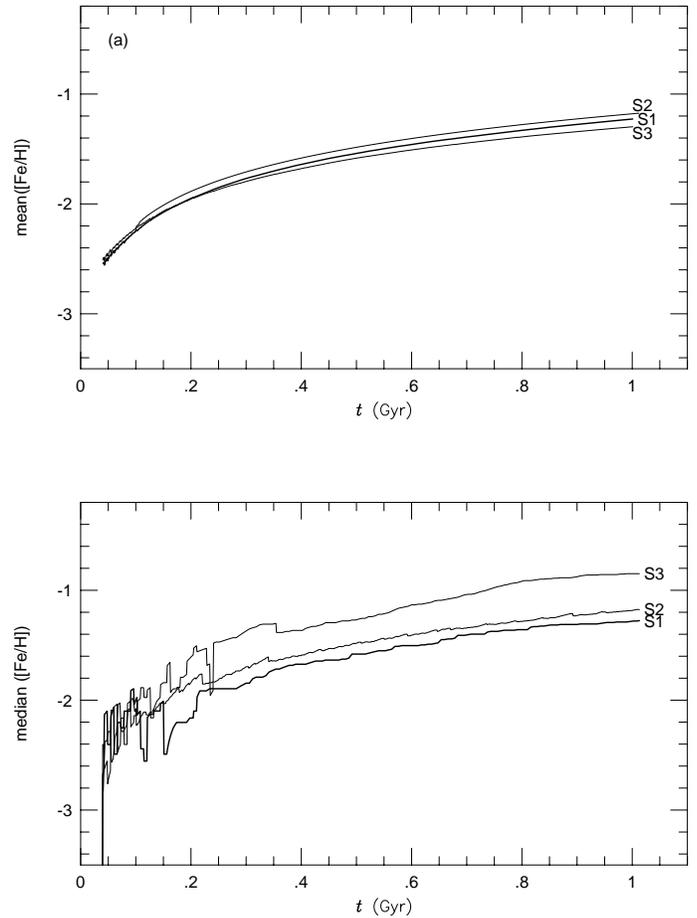} 
\caption[]{
(a)~~The evolution of the mean iron abundance 
in the standard models\,S1, S2, and S3. 
(b)~~
The same as figure\,2(a), but for the median iron abundance. 
}
\end{center}
\end{figure}

Figure~3 illustrates a frequency distribution of iron abundance  
of the 1000 blocks, $N$([Fe/H]), of model\,S1 at 0.01, 0.05, 0.1, 0.2, 0.5, 
and 1\,Gyr. For an illustrating purpose, $N$($-5 <$[Fe/H]$\le -4.8$) 
is artificially assigned for 
a fraction of metal-free blocks. Figure~3 shows that the metallicity 
distribution of model\,S1 has roughly 1.3 \,dex spread of the iron 
abundance, suggesting that the ISM in the halo was inhomogeneous till 1\,Gyr. 

\begin{figure*}[t]
\begin{center}
\epsfxsize=9cm
\epsfbox{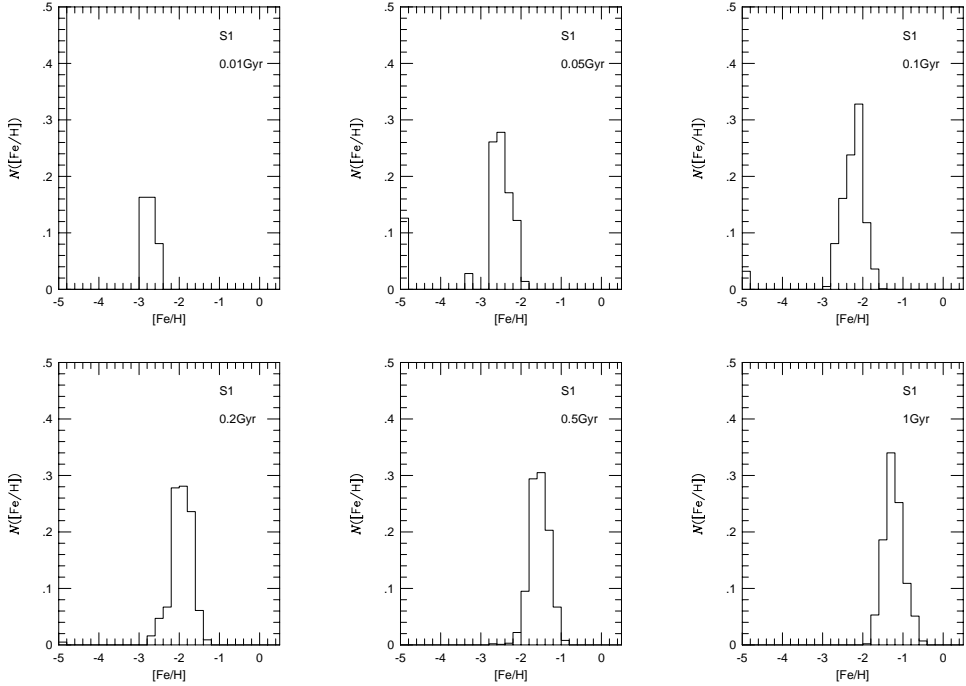}
\caption[]{
The frequency distribution of iron abundance of the 1000 blocks in model\,S1 
at 0.01, 0.05, 0.1, 0.2, 0.5, and 1\,Gyr. 
}
\end{center}
\end{figure*}
\begin{figure}[h]
\epsfxsize=5cm
\epsfbox{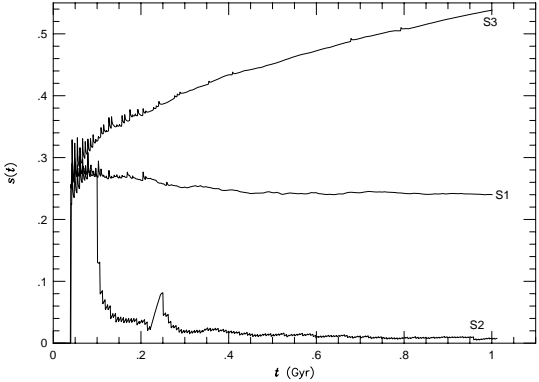}
\caption[]{
The evolution of standard deviation of metallicity distribution 
in the standard models\,S1, S2, and S3. 
}
\end{figure}

Figure\,4 gives the evolution of standard deviation of the metallicity 
distribution obtained by the standard models. In model\,S1, the standard 
deviation keeps nearly constant at $s(t) \simeq 0.24$ after $\sim 0.2$\,Gyr, 
which corresponds to roughly $1.3$\,dex dispersion of the iron 
abundance. Model\,S3 suggests much stronger abundance spread than model\,S1. 
The standard deviation of model\,S3 increases with time and exceeds 
$0.5$ at 1\,Gyr, which is roughly equivalent to 2~dex dispersion of the iron 
abundance among all the enriched blocks. 
On the other hand, the standard deviation of model\,S2 decreases  
quickly to nearly zero, showing that the ISM was homogeneous 
from the very beginning of its chemical evolution.

\begin{center}
\begin{figure}
\epsfysize=10cm
\epsfbox{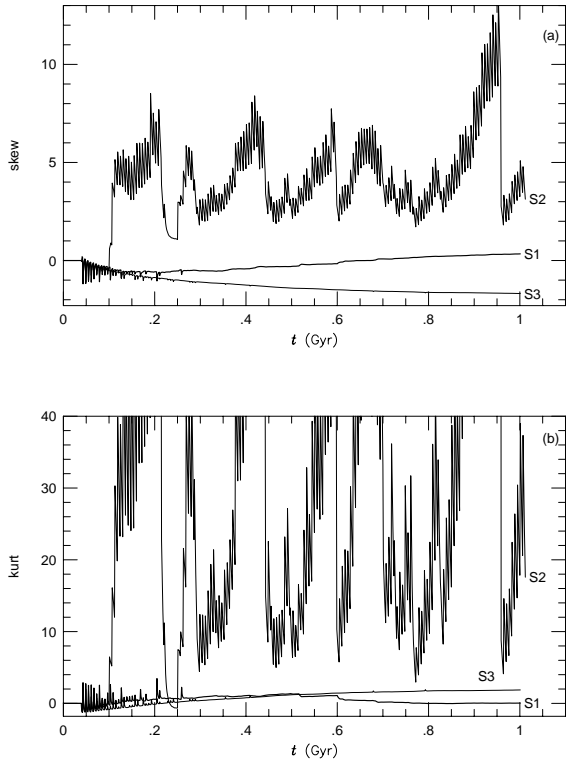}
\caption[]{
(a)~~The evolution of skewness of metallicity distribution 
of the standard models\,S1, S2\, and S3. 
(b)~~The same as figure\,5a, but for the kurtosis 
of metallicity distribution. 
}
\end{figure}
\end{center}
Now we focus on the shape of the metallicity distribution. 
Figures~5\,a~and~5\,b illustrate the skewness and the kurtosis of the 
metallicity distributions of the standard models, respectively 
(e.g. Stuart \& Ord 1987). The skewness characterises a degree of asymmetry 
of a distribution around its mean. 
It characterises only a shape of the distribution. The definition is 
\begin{equation}
{\rm {Skew}}(x_1,~x_2,....,~x_N)= \frac{1}{N} \sum_{i=1}^{N}(\frac{x_i-\bar{x}}{s})^3,
	\label{eqn:skew}
\end{equation}
where $\bar{x}$ and $s$ are the mean and the standard deviation 
of the measured value $x_1,~x_2,...,~x_N$. A positive value of skewness 
signifies a distribution with an asymmetric tail extending out towards 
larger $x$, while a negative value signifies a distribution whose tail 
extends out towards smaller $x$. 
The definition of the kurtosis is
\begin{equation}
{\rm{Kurt}}(x_1,~x_2,....,~x_N)= \{ \frac{1}{N}
 \sum_{i=1}^{N}(\frac{x_i-\bar{x}}{s})^4 \} -3,  
	\label{eqn:kurt}
\end{equation}
where the first term of r.h.s. becomes 3 for a Gaussian distribution. 
The kurtosis measures the relative peakedness 
or flatness of a distribution relative to the Gaussian distribution. 
The distribution with positive kurtosis has the outline of the Matterhorn 
for example. The distribution with negative kurtosis is outlined 
of a lump of meat-loaf. 
Figure \,5a gives the evolution of skewness of the standard models. 
The skewness of model\,S1 is nearly zero, which means a symmetric 
shape of the metallicity distribution. The skewness of model\,S2 behaves 
irregularly due to nearly null standard deviation.
The skewness of model\,S3 is always negative and decreases
as times goes on, 
showing a tail of the metallicity distribution extending toward lower 
metallicity at later stages. This is due to the localised chemical enrichment 
caused by the 
stimulated star formation. Figure\,5b shows the evolution of kurtosis. 
The kurtosis of model\,S1 is nearly equal to zero, thus 
the metallicity distribution has a Gaussian like shape.
On the contrary, the kurtosis of 
model\,S2 is very large. This model suggests the homogeneous ISM, 
since the kurtosis should be infinitely large if samples have the same value. 
In model\,S3, the kurtosis gradually increases up to $\sim$ 2, 
which means that 
the metallicity distribution has a sharp peak. Thus the skewness and 
the kurtosis of model\,S3 indicate that the metallicity distribution 
has a peaked shape with a tail extending to the lower metallicity. 

As a summary, only model\,S2 shows that the ISM was chemically 
homogeneous in the early halo, while both models\,S1~and~S3 show 
that the ISM in the halo was not well-mixed till at least 1\,Gyr. Indeed, 
the scatters of the iron abundance are roughly
1.4 dex in model\,S1 and 2 dex in model\,S3, respectively. 
We have repeated the simulations several times  and confirm 
that the model results show always the same tendencies. 

\section{Observational Constraints \label{subsubsec:gp}}

Long-lived stars in the halo should keep the original metallicities
of the ISM from which they formed. It is true that little observational
information is available for constraining chemical evolution models
of the halo, but we will show that a cumulative metallicity 
distribution function of the
long-lived halo stars can potentially give a powerful clue to obtain the
best model prescriptions. 

\subsection{Comparison with the Standard Models}

Figures\,6~and~7 present generalised metallicity 
distribution functions (GMDFs), $P$([Fe/H]), of the standard models
S1 -- S3 together with the empirical GMDFs of long-lived halo stars
taken from Ryan \& Norris (1991) and Carney et al. (1996). 
\begin{figure*}[t]
\begin{center}
\epsfxsize=9cm
\epsfbox{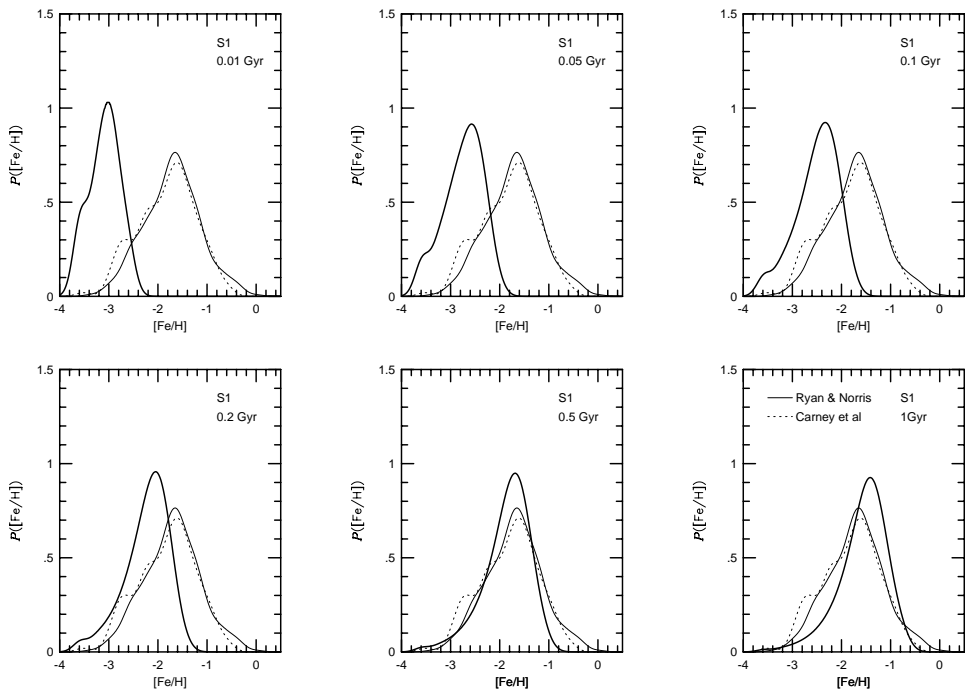}
\caption[]
{
The generalised metallicity distribution functions (GMDFs) of
model\,S1 at 0.01, 0.05, 0.1, 0.2, 0.5, and 1\,Gyr (thick solid line).  
Thin solid and dotted lines illustrate the empirical halo GMDFs taken from 
Ryan \& Norris (1991) and Carney et al. (1996), respectively.
}
\end{center}
\end{figure*}
The definition of GMDF is given by Laird et al. (1988):  
\begin{equation}
P(x)=\frac{1}{N \sigma \sqrt{2 \pi}} 
\sum_{i=1}^{N} \exp\left [-\frac{(x-x_i)^2}{2 \sigma^2} \right ], 
\end{equation}
where $\sigma$ reads as a typical error in the observed values
$x_1$, $x_2$, ..., $x_N$. 

Quite often, the empirical metallicity distribution function is presented 
in the form of a histogram. However, Searle \& Zinn (1978) suggested that 
the bins of a conventional histogram should be replaced by a continuous 
distribution function such as a Gaussian to obtain a better approximation 
to the actual abundance distribution function, since the binning distorts 
the data. Therefore, we convolve the observed histograms of stellar
metallicity distribution to include the uncertainty of observational data 
properly (Laird et al 1988) and 
convolve the theoretical GMDFs by using equation (16) 
with $\sigma=0.15$ (Ryan \& Norris 1991; Carney et al. 1996).
We note that the GMDFs of Ryan \& Norris (1991)
and Carney et al. (1996) are almost identical, except that the latter shows
a small hump at [Fe/H] $ \simeq -2.7$.

Figure\,6 shows that model\,S1 gives good fits to the observed GMDFs   
at 0.5~Gyr and 1~Gyr. The star formation in the 
halo must have virtually stopped at around 0.5 Gyr, otherwise the resulting
GMDF gives too high peak metallicity. This would happen if the gas escapes
from the halo and accretes on to the bulge and disc dissipationally.
A precise value of the epoch when the star formation terminated is 
of course model dependent. However, the GMDFs of
halo stars strongly suggest that the star formation in the halo did not
last longer than, say, 0.5-1 Gyr.

Upper and lower panels in figure\,7 illustrate the GMDFs of models\,S2~and~S3, 
respectively. Clearly, model\,S2 is inconsistent with the observations. 
Model\,S2 gives too sharp GMDFs to reproduce the observations.
We have also studied several alternative cases in which we assume much 
longer time interval for the periodic turbulent mixing, much smaller
size of the smoothed out area, and lower or higher probability 
of star formation, and have confirmed that these models give
much sharper GMDFs than the empirical ones; thus inconsistent 
with the observations. On the contrary, the GMDF of 
model\,S3 is consistent with those observed. 
Equivalent models to model\,S3, but with higher or lower SFRs,
are also calculated. These models predict similar inhomogeneous ISM 
enrichment. Their GMDFs are also consistent with the empirical ones 
at 0.1\,Gyr (higher SFR) and 0.5\,Gyr (lower SFR).  
The standard deviations of metallicity distribution function of these models 
increase with time, showing the same trends as those of model\,S3. 
Without additional informations, 
it is rather difficult to conclude which star formation
mechanism, the spontaneous star formation (model\,S1) 
or the stimulated one (model\,S3),
is responsible for the early enrichment of the Galactic halo.

\begin{figure*}[t]
\epsfxsize=9cm
\epsfbox{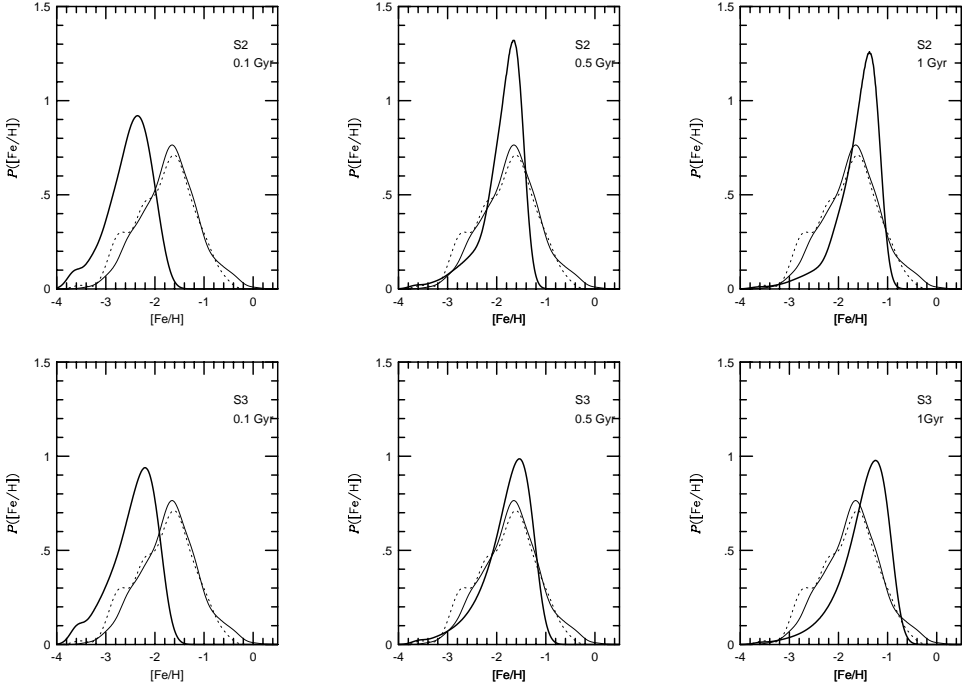}
\caption[]
{
The GMDFs of model\,S2 (upper three panels) and 
model\,S3 (lower three panels) at 0.1, 0.5, and 1\,Gyr.  
Thick solid lines show the theoretical GMDFs. 
Thin solid and dotted lines have the same meaning as in Fig.\, 6. 
}
\end{figure*}

To judge a goodness of the model fit to the empirical GMDFs,    
we have performed the $\chi^2$ statistics. 
Let $x_i$ and $y_i$ be the theoretical and the observed  $P$([Fe/H]$_i$),
respectively. Then, the $\chi^2$ value is defined as, 
\begin{equation}
\chi^2=\sum_i \frac{(x_i-y_i)^2}{x_i+y_i}. 
\end{equation}
In columns (8)-(10) of table 2, we give the $\chi^2$ values of
the standard models at 0.1, 0.5, and 1\,Gyr, respectively. The best fit
is realised by model S1 at 0.5 Gyr, but model S1 at 1 Gyr and model S3
at 0.5 Gyr also give a reasonable fit.

\begin{table*}
\small
\begin{center}
Table~2.\hspace{4pt}Models and results.\\
\vspace{6pt}
\end{center}
\vspace{6pt}
\begin{tabular*}{\textwidth}{@{\hspace{\tabcolsep}
\extracolsep{\fill}}lrrlcrrrrrr}
\hline\hline\\[-6pt]
Model & $l_{\rm b}/R_{\rm h}$ & {\scriptsize $\omega^{-1} \Delta t$} & & & \multicolumn{3}{c}{$s(t)$} & \multicolumn{3}{c}{$\chi^2$}  \\
\multicolumn{3}{c}{} &  & & 0.1 &0.5 &1 & 0.1 &0.5 &1 \\  
\hline
S1 & 0.03 & 1.2 &~~& & 0.27 & 0.24 & 0.24 &~~ 75.11 & 11.01 & 11.85 \\
S2 & 0.03 & 1.2 &  & & 0.27 & 0.01 & 0.01 & 76.82 & 26.78 & 22.63 \\
S3 & 0.03 & 1.2 &  &~~& 0.33 & 0.45 & 0.54 & ~~68.16 & 11.89 & 20.36 \\
& & \\
A & 0.03 & 0.6 &   &~~& 0.34 & 0.34 & 0.29 & ~~53.52 & 4.68 & 15.82 \\
B & 0.03 & 3 &  &~~& 0.18 & 0.10 & 0.10 & ~~95.12 & 24.40  & 16.79 \\
C & 0.01 & 1.2 &  &~~& 0.32 & 0.44 & 0.42 & ~~9.25 &  11.68 & 19.73 \\
D & 0.03 & 1.2 & $n_{\rm{II}}=20$  &~~& 0.15 & 0.11 & 0.11 & ~~92.31 & 21.39 & 17.19 \\
E & 0.03 & 1.2 & $m_{\rm{II}}$=8$M_\odot$  &~~& 0.24 & 0.16 & 0.17 & ~~83.33 & 16.70 & 14.02 \\
F & 0.03 & 1.2 & $m_l=0.05M_\odot$  &~~& 0.30 & 0.33 & 0.30 & ~~62.80 & 6.19 & 11.77 \\
G & 0.03 & 1.2 &  $M_{\rm h}=2 \cdot 10^{11}M_\odot$ & & 0.30 &  0.42 & 0.37 & ~~49.38 & 4.00 &  13.06 \\
\hline
\end{tabular*}

\vspace{6pt}

\noindent
Model\,S2 assumes propagation of star formation. Model\,S3 assumes  
the mixing of ISM by turbulent diffusion whose priod is taken 
to be $10^8$~yr. See text in detail.

\end{table*}

In addition to the standard models, we show additional 7 models A - G 
to demonstrate the parameter dependence of our results. 
All the 7 models are equivalent to model\,S1, but with different parameters, 
i.e., these models assume spontaneous star formation and no turbulent mixing. 
The parameters of these 7 models are summarised in table 2.
In this table, column (1) gives an identification of the model, columns 
(2) and (3) give the adopted values of 
$l_{\rm b}/R_{\rm h}$ and $\omega^{-1} \Delta t$, respectively.
Column (4) gives the parameter that is different from the standard models. 
Columns (5) - (7) indicate the standard deviation $s(t)$ at
0.1, 0.5, and 1\,Gyr, respectively.
The $\chi^2$ values for the best fit epochs of the standard models 
are 11.01 (model\,S1) and 11.89 (model\,S3) at 0.5\,Gyr. 
Therefore, we consider a good fit is realised if the model gives the 
$\chi^2$ less than 12 in columns (8) - (10). 

In models A - C,  we study how the evolution of $s(t)$ and $\chi^2$
depend on $l_{\rm b}/R_{\rm h}$ and $\omega^{-1} \Delta t$.
Two cases of $l_{\rm b}/R_{\rm h}=$ 0.01 and 0.03 are studied.    
For $\omega^{-1} \Delta t$, we consider three cases, 
i.e. $\omega^{-1} \Delta t=0.6 \cdot 10^{-3}$, 
$1.2 \cdot 10^{-3}$, and $3 \cdot 10^{-3}$.    
The standard deviation of model\,A is always larger than
that of model\,S1, which suggests that the ISM is always 
chemically inhomogeneous. In model\,B, higher SFRs lessen  
the differences of chemical enrichment among the blocks. 
The standard deviation 
of model\,B is nearly equal to $0.1$ at 1\,Gyr, predicting relatively 
well-mixed ISM with respect to model\,S1. However, the $\chi^2$  
of model\,B indicates that this model is inconsistent with the observations.  
In model\,C, given smaller $l_{\rm b}/R_{\rm h}$, the best fit epoch 
to the observations comes earlier than model\,S1. 
The gas in the star forming blocks of model\,C is converted into stars more 
efficiently than model\,S1, since  
equation (14) indicates that the SFR in a block is proportional to 
$n_{\rm{II}}/m_{\rm b}$, i.e., $ (l_{\rm b}/R_{\rm h})^{-3}$. 
In model\,D, we reduce the number of SN\,II progenitor 
in a single OB association and assume $n_{\rm{II}}=20$  
instead of 40 (Blaauw 1964). This 
increases the probability of star formation (see equation (11)) and 
homogenises the ISM quickly. 
The $\chi^2$ of model\,D indicates poor fits to the observations.  
Model\,E assumes $m_{\rm{II}}=8M_\odot$. 
The number fraction of SN\,II progenitor in model\,E is increased 
and the SFR becomes twice of model\,S1. 
Thus, both the evolutionary behaviour of standard deviation 
and the $\chi^2$ fits are very similar to those of model\,B. 
We can reject this model, since the $\chi^2$ gives very poor fits.
Models\,F~and~G assume $m_l=0.05M_\odot$ 
and $M_{\rm h}=2 \cdot 10^{11}\,M_\odot$, respectively. Equation (11) 
indicates that the probability of star formation $P(t)$ in a block, 
namely the SFR in the halo, is proportional to the number 
fraction $f_{\rm{II}}$ of SN\,II progenitor 
and the mass of the halo. Thus 
in models\,F~and~G with the lower SFR in the halo, 
the standard deviation of metallicity distribution 
function becomes slightly larger than that of model\,S1. 
The $\chi^2$ values of models\,F~and~G show that these models 
give a good fit to the observed GMDFs in the halo at 0.5\,Gyr. 

In summary, the different parameters change the SFR in the halo.  
If the SFR is low, the standard deviation of metallicity 
distribution becomes large, i.e. the ISM is inhomogeneous, and 
good fits to the observed GMDFs are resulted. 
While if the SFR in the halo is high, 
the standard deviation becomes small (well-mixed ISM) and  
the resulting GMDFs show poor fits to the observations, since  
stars born in the similarly enriched ISM dominate 
the cumulative metallicity distribution function at later phase. 
Thus the predicted GMDF becomes too narrow to be consistent with 
the observations. 

\subsection{Comparison with Mass-Loss Models}
In previous sections, we assume that the halo was a closed system. 
Hartwick (1976) showed that a simple model predicts 
too many metal-poor stars and is inconsistent with 
the cumulative distribution of the iron abundances 
for metal-poor ([Fe/H]$\le -0.52$) globular clusters in the Galactic halo. 
He modified the simple model 
and proposed a model which allows for gas outflow from the halo. 
We therefore consider effects of mass loss from the halo in this subsection. 
The gas will eventually either escape to intergalacitc space or to accrete 
onto the galactic plane. 

Since the gas removal from the halo could be related to energy injection 
from hot OB stars and SN\,II (Hartwick 1976), we assume that the mass loss 
rate $dD/dt$ is proportional to the SFR $C(t)$:
\begin{equation}
\frac{dD}{dt}=b\,C(t). 
\end{equation}
The mass loss rate $b$ and the SFR are taken from 
Hartwick (1976), i.e., $b=10$ and $\omega^{-1}=0.3$, respectively. 
We assume the same gas outflow rate for all the blocks in 
the halo. In model ML1, the parameters are the same as those of model S1 
except for $b$ and $\omega^{-1}$. We assume that OB associations are born 
randomly in space and do not consider the ISM mixing by turbulent 
diffusion. While in model ML2, we study the effects of turbulent mixing. 
We assume periodic ISM mixing, similarly to model S2. The period is 
taken as 0.1\,Gyr, the same as that of model S2. Other prescriptions 
for model ML2 are same as those of model ML1. 

Figures\,8a~and~8b show the evolution of mean iron abundance and 
that of standard deviation of the mass loss models, respectively.
Figure~9  shows the theoretical GMDFs plotted with the observed GMDFs.   
These figures show that the GMDFs obtained by model\,ML1 agree with 
the observed GMDFs and the ISM is always 
inhomogeneous in this model. On the other hand, 
the GMDFs of model\,ML2 are too sharp and 
are inconsistent with the observations, 
because the ISM in this model is well-mixed. 
In model ML1, the best fit is achieved at an epoch later than model S1. 
In the case of one-zone model, a shift of the peak abundance can be 
explained by a decrease of effective yield caused by the mass-loss 
(Hartwick 1976; Pagel 1992). Although our model is not one-zone, 
the mass-loss gives a similar effect to the peak abundance of 
the cumulative metallicity distribution function.  
 
\begin{figure}[t]
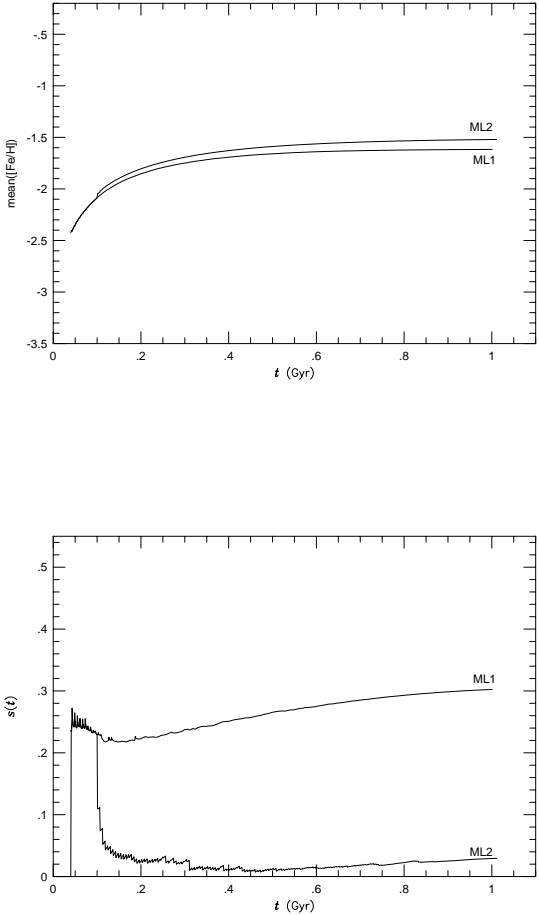

\epsfxsize=5cm
\epsfbox{fig8a.epsi}
\epsfxsize=5cm
\epsfbox{fig8b.epsi}
\caption[]{
(a) ~~
The same as figure\,2(a), but for mass loss models ML1 and ML2. 
The SFR $\omega^{-1}=0.3$ and the mass-loss rate $b=10$ are the same as those 
of Hartwick (1976). 
(b) ~~
The same as figure\,4, but for mass-loss models\,ML1 and ML2.
}
\end{figure}
\begin{figure*}[ht]
\epsfxsize=9cm
\epsfbox{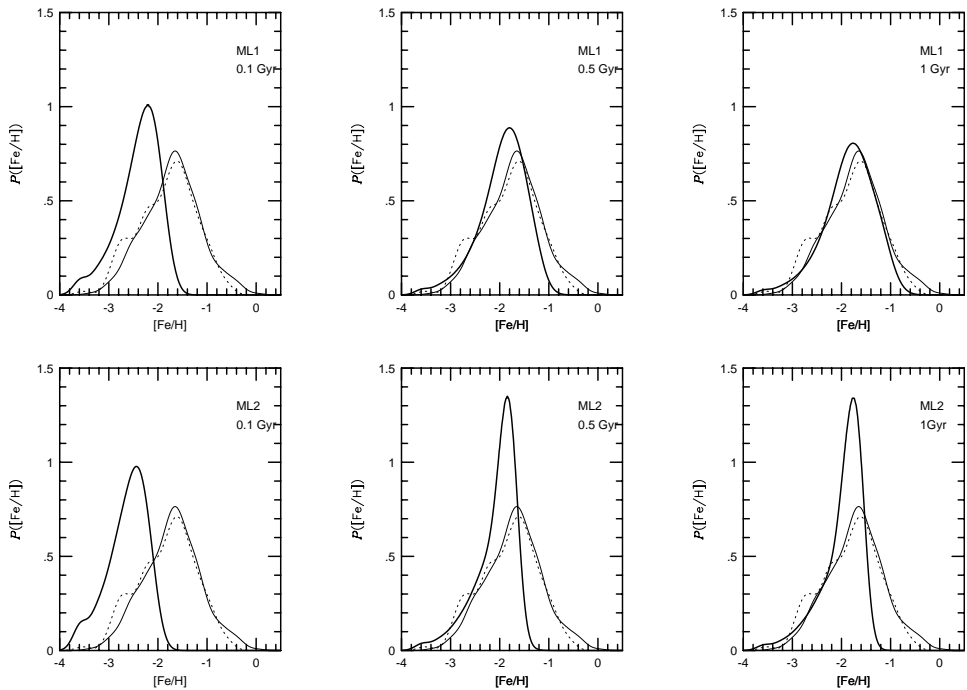}
\caption[]{
The GMDFs of model\,ML1 (upper three panels) and 
model\,ML2 (lower three panels) at 0.1, 0.5, and 1\,Gyr.  
Thick solid lines show the theoretical GMDFs. 
Thin solid and dotted lines have the same meaning as in Fig.\, 6. 
}
\end{figure*}

The GMDFs obtained by assuming $b=10$ and $\omega^{-1}=0.2$ 
(the same SFR as that of model~S1) also fit to the observed GMDFs and 
the best fit is realised at around 1\,Gyr. 
The chemical enrichment in this model 
is also inhomogeneous.  
A model adopting $b=5$ also gives consistent GMDFs with 
the observations, although 
the best fit epoch appears earlier than model\,ML1 due to a larger effective  
yield. Gas outflow may occur in the vicinity of star forming regions, 
if major energy sources are massive 
stars. When we take this into account and consider a model 
in which gas is expelled only from the star forming blocks, 
we obtain the same results on a fitting goodness and 
inhomogeneity of the ISM as those obtained by models\,ML1 and ML2.

As a conclusion, whether or not mass loss is taken into account, 
the ISM in the halo at the beginning must be 
inhomogeneous. 

\section{Discussions \label{sec:discussion}}

Observations show neither clear age-metallicity relation nor clear 
evidence for metallicity gradient in the halo. 
Because of these observational results,
Galactic halo formation scenario favoured recently 
(e.g Carney et al. 1996) is  
the one proposed by Searle \& Zinn (1978), i.e. 
accretion of small proto-galactic fragments contributed to the 
halo population. However, in the poorly mixed halo, 
the age-metallicity relation and the metallicity gradient 
cannot be expected even if the Galaxy formed from one massive cloud. 
Thus it is not necessary to introduce the accretion 
of proto-galactic fragments 
to interpret the lack of the age-metallicity relation 
and the metallicity gradient in the halo. 

The scatters observed in the relative abundances of neutron capture elements 
to the iron, particularly [Sr/Fe], are often claimed 
as evidences for the inhomogeneous enrichment of the ISM 
in the halo at the very beginning. 
In figure\,10, Sr abundances of halo stars taken from 
Gratton \& Sneden (1988; 1994), Magain (1989), 
McWilliam et al. (1995), and Ryan et al. (1996) are plotted. 
The observations show that a majority of the metal-poor stars 
with [Fe/H]\,$\le -2.5$ distribute at $-1.5 \le$\,[Sr/Fe]\,$\le 0.6$. 
The relative abundance of Sr tends to decrease clearly with an increase of 
the iron abundance. 

We show two theoretical evolutionary paths of Sr on figure\,10. 
The models are calculated by adopting  
the same parameters as those of models\,S1~(left path)~and~C (right path). 
The SFR of model\,S1 is nearly one tenth of model\,C.   
Co-existence of regions with different SFR is a view of 
the chemical evolution in the halo considered here. We have simplified 
the situation and assumed the same SFR in each block. 
However, different SFR in each star forming region is more realistic. 
These models are consistent with the observed behaviour of [Sr/Fe] 
and the empirical metallicity distribution functions.

Following PT95 and Pagel \& Tautvai$\check{\rm{s}}$ien$\dot{\rm{e}}$ 
(1997; hereafter PT97)),  
we calculate the chemical evolution of Sr: 
\begin{equation}
\frac{d}{dt}(gZ_{\rm{Sr}})=-Z_{\rm{Sr}}(t)c(t)
+p_0c(t)+ \sum_{i=1}^{3} p_ic(t-\tau_i), 
\end{equation}
where $g(t)$, $c(t)$, and $Z_{\rm{Sr}}(t)$ are the gas fraction, 
the SFR,  and abundance of Sr in a block, respectively. 
The first and the second terms 
on the r.h.s. have the same meaning as those in equation\,(12). 
The third term comes from the delayed production,
where $p_i$ is the yield corresponding to 
the fixed time lag $\tau_i$. Here $p_0=0.01$, $p_1=0.08$, 
$p_2=0.39$, $p_3=0.23$, $\tau_1=0.023$\,Gyr, 
$\tau_2=0.025$\,Gyr, and $\tau_3=2.7$\,Gyr are assumed (PT97).      
The evolution of the iron abundance is calculated by equation\,(12). 

The theoretical paths in figure\,10 show that [Sr/Fe] begins 
to increase rapidly 
at [Fe/H]\,$\sim -3.4$ in model\,S1 and [Fe/H]\,$\sim -2.3$ in model\,C,  
depending on the SFR. These paths roughly outline the dispersion of 
observed [Sr/Fe] at [Fe/H]\,$\le -2.5$. 
The models are calculated with a time step of $0.001$\,Gyr, 
which is shown by filled pentagons on the theoretical paths in figure\,10. 
The chemical evolution from [Sr/Fe]\,$\sim -1.3$ to [Sr/Fe]\,$\sim 0$ 
should take place within a short time interval ($\sim 0.012$\,Gyr).
After the rapid increase, 
the paths show temporal increase and decrease, and then converge. 
The declines of theoretical [Sr/Fe] are not because of 
delayed production of the iron (in other words, 
contribution from SNIa which produces 
the bulk of iron in the solar neighbourhood), since  
we assume that the time delay of iron $\tau$ 
in equation (\ref{eqn:delay}) is equal to $1.3$\,Gyr (see section 2.3) 
and consider chemical evolution before $1$\,Gyr. 
The declines of [Sr/Fe] are due to 
onsets of the next formation of OB associations.  
Thus different SFR in each star forming region can explain 
the trends defined by a majority of observed stars 
on [Sr/Fe] vs [Fe/H] diagram.  

\begin{figure}[t]
\epsfxsize=6cm
\epsfbox{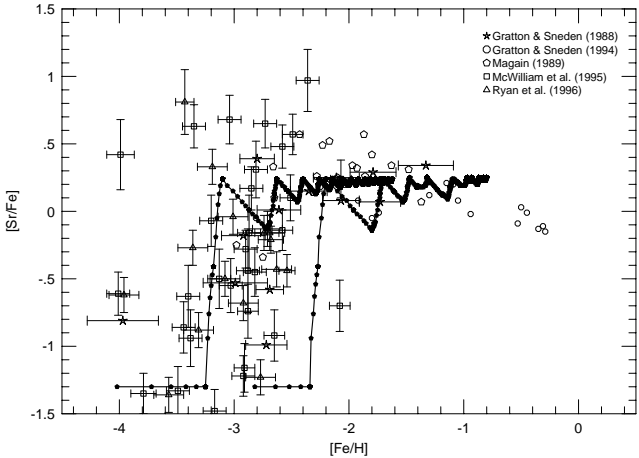}
\caption[]{
Chemical evolution of strontium.  Observational data are taken  
from Gratton \& Sneden (1988; 1994), Magain (1989), 
McWilliam et al. (1995), and Ryan et al. (1996). 
Two theoretical paths correspond to different SFR in a block. 
The left and right paths are obtained by adopting the same parameters 
as those of models\,S1~and~C, respectively. 
Each filled pentagon on the paths indicates 
a time step $0.001$~Gyr. 
}
\end{figure}

Apparently the models are inconsistent with the observational 
data on figure\,10. 
McWillam et al (1995), Sneden et al (1996), and McWillam (1998) 
have shown that the heavy element abundances in the super-solar 
[Sr/Fe] stars are dominated by the r-process abundance patterns.  
The r-process elements can only be produced by Type II supernova events. 
Therefore one may argue that the 
time-delay model for Sr is inadequate to explain the observations. 
If the strontium is produced in all progenitors of SN\,II, however, 
[Sr/Fe] values observed in metal-poor stars should be always super-solar. 
On the contrary, as we mentioned before, figure 10 shows that [Sr/Fe] 
tends to decrease at lower iron abundance ([Fe/H]\,$\le -2.5$). 
This drop suggests that the strontium was formed slightly later 
than the iron (see Mathews, Bazan, \& Cowan 1992). 

If the large scatter of [Sr/Fe] observed for stars of iron abundance in 
the range of $-3.6 \le$[Fe/H]$\le -2.6$ is indeed due to a sharp increase 
of the Sr production rate, the Sr abundances of four extremely metal-dificient 
stars with [Fe/H]$\sim -4$ seem to show some evidence against it. However, 
the iron abundance measurement of such very low metal stars is 
extremely difficult and these iron abundances could well be as large as 
[Fe/H]$\stackrel{>}{_\sim}-3.6$ (M. Spite, private communications). 

The theoretical [Sr/Fe] is too low to fit to the observed 
extremely high [Sr/Fe] value. 
However, the abundance of these stars might not reflect 
the composition of ISM from which they formed (e.g. McClure 1984). 
We should reject CH stars when we discuss the Galactic chemical evolution, 
because atomosphers of CH stars are thought to have been enriched 
by elements transfered from evolved companion AGB stars (McClure 1984). 
McWillam et al (1995) and McWilliam (1998) reported that 
CS22898-027 ([Fe/H]$=-2.36$) and CS22947-187 ([Fe/H]$=-2.5$) are 
probably CH stars. 
Observational errors might also reflect the [Sr/Fe]-[Fe/H] diagram.   
For a star CS22891-209 ([Fe/H]$\sim -3.2$),  
Primas, Molaro, \& Castelli (1994) and McWilliam et al (1995) reported [Sr/Fe]$=0.92$ and 
[Sr/Fe]$=-0.07$, respectively, although 
they used the same lines and had the same S/N.

\section{Conclusion}  

A stochastic chemical evolution model has been built to study an early 
history of metal enrichment in the Galactic halo. The metallicity 
distribution function of long-lived halo stars is found to be a 
clue to obtain the best model prescriptions. 
We find that the star formation in the halo virtually terminated by 
$\sim 1$ Gyr and that the halo has never been chemically homogeneous 
in its star formation history. The star formation in the halo 
could either be spontaneous or stimulated, which keep the halo always 
inhomogeneous and the turbulent mixing is found to be inefficient. 
This conclusion does not depend whether the mass loss from the halo is taken 
into account or not. The observed ratios of the $\alpha$-elements with 
respect to the iron do not show scatters on the [$\alpha$/Fe]-[Fe/H] 
plane, but this does not imply that the ISM in the halo was homogeneous 
because the chemical evolution path on this diagram is degenerate in the 
SFR. On the other hand, the apparent spread of [Sr/Fe] ratio among 
metal-poor halo stars does not reflect an inhomogeneous metal enrichment, 
instead it is due to a sharp increase in the production rate of 
strontium that is probably synthesised in slightly less massive stars 
than the progenitor of iron-producing SN\,II.

\vspace{1cm}

We are grateful to an anonymous referee for a careful reading 
of the manuscript and for useful comments.
C.I. thanks to the Japan Society for Promotion of 
Science for a financial support. 
This work was financially supported in part 
by a Grant-in-Aid for the Scientific Research (No. 0940311) 
by the Japanese Ministry of Education, Culture, Sports
and Science. 


\section*{References}
\small

\re
Allen D.A., Burton, M.G. 1993, Nature 363, 54
\re Arimoto N., Matsushita, K., Ishimaru, Y., Ohashi, T., Renzini A. 1997, ApJ 477, 128
\re Arimoto N., Yoshii, Y., Takahara, F. 1992, A\&A 253, 21
\re  Audouze J., Silk J. 1995, ApJ 451, L49 
\re Bahcall J.N., Pinsonneault M.H. 1996, AAS 189, 5601
\re  Bateman N.P.T., Larson R. 1993, ApJ 407, 634
\re Barbuy B. 1988, A\&A 191, 121
\re  Beers T.C., Preston G.W., Shectman S.A. 1992, AJ 103, 1987
\re  Blaauw A. 1964,  ARA\&A 2, 213
\re  Bruhweiler F.C., Gull, T., Kafatos M.,  Sofia S. 1980, ApJ 238, L27
\re Cash W., Charles P., Bowyer S., Walter F., Garmire G., Riegler G. 1980, ApJ 238, L71
\re Carraro G.,  Chiosi C. 1994, A\&A 281, 35
\re Carney S.G., Laird J.B., Latham D.W., Aguilar L.A. 1996, A\&A 112, 668 
\re  Cioffi D.F., McKee C.F., Bertschinger E. 1988, ApJ 334, 252
\re Copi C.J. 1997, ApJ 487, 704
\re  Cox D. P. 1972, ApJ 178, 159
\re  Edvardsson B., Andersen J., Gustafsson B., Lmbert D.L., Tomkin J. 1993, A\&A 275, 101
\re  Elmegreen B.G. 1982, in Submillimeter Wave Astronomy, 
ed. J.E. Beckman, J.P. Phillips (Cambridge University Press) p5
\re  -----. 1985a, in Birth and Infancy of Stars, 
ed. R. Lucas, A. Omont,  R. Stora (Amsterdam: Elsevier) p215 
\re  -----. 1985b, in Brith and Evolution of Massive Stars 
and Stellar Collapse, ed W. Boland, 
H. van Woerden (Dordrecht: Reidel) p227
\re  Elmegreen B.G., Lada, C.J. 1977, ApJ 214, 725
\re  Fich M., Tremaine, S. 1991, ARA\&A 29, 409
\re  Gratton R.G., Sneden C. 1988,  A\&A 204, 193
\re  Gratton R.G., Sneden C. 1994,  A\&A 287, 927
\re  Hartwick, F.D.A., 1976, ApJ 209, 418
\re  Henning T., G$\ddot{\rm u}$rler J. 1986, Ap\&SS 128, 199
\re  Heiles C. 1987, ApJ 315, 555
\re  van den Hoek L.B., de Jong T. 1997, A\&A 318, 231 
\re  Hoyle F., 1953, ApJ 118, 513
\re Humphreys, R.M. 1978, ApJS, 38, 309
\re  Lada C.J., Blitz L., Elmegreen B. 1979, in Protostars and Planets, ed T. Gehrels (Tucson: University of Arizona Press) p368 
\re  Laird J.B., Rupen M.P., Carney B.W., Latham D.W. 1988, AJ 96, 1908
\re Larson R.B. 1974 MNRAS, 169, 229
\re  Mac Low M.-M., McCray R. 1988, ApJ 324, 776
\re  Magain P. 1989, A\&A  209, 211
\re Mathews, G.J., Bazan, G., Cowan, J.J. 1992, ApJ 391, 719
\re  McCray R., Kafatos M. 1987, 317, 190
\re McClure R.D. 1984, PASP, 96, 117 
\re  McWilliam A 1997, ARA\&A 35, 503
\re  McWilliam A 1998, AJ 115, 1640
\re  McWilliam A., Preston G. W., Sneden C., Shectman S. 1995, AJ 109, 2757
\re Nissen P.E., Gustafsson B., Edvardsson B., Gilmore G. 1994, A\&A 285, 440
\re  Pagel B.E.J., 1992, in The Stellar Populations of Galaxies, IAU Symposium No. 149, edited by B. Barbuy \& Renzini (Kluwer, Dordrecht), p. 133 
\re  Pagel B.E.J., Tautvai$\check{{\mbox s}}$ien$\dot{{\mbox e}}$ G. 1995, MNRAS 276, 505 (PT95)
\re  Pagel B.E.J., Tautvai$\check{{\mbox s}}$ien$\dot{{\mbox e}}$ G. 1997, MNRAS 288, 108 (PT97)
\re  Pilyugin L.S. 1996, A\&A 313, 803
\re Primas F., Molaro, P., Castelli, F. 1994, A\&A 290, 885
\re  Roy J.-R., Kunth D. 1995, A\&A 294, 432
\re  Ryan S.G., Norris J.E. 1991, AJ 101, 1865
\re  Ryan S.G., Norris J.E., Beers T.C. 1996, ApJ 471, 254
\re Saito M. 1979, PASJ 31, 181
\re  Salpeter E.E. 1955, ApJ 121, 161
\re Searle L., Zinn R. 1978, ApJ 225, 357
\re  Schmidt M 1959, ApJ 129, 243
\re  Seal L., Zinn R. 1978, ApJ 225, 357
\re  Silk J. 1977, ApJ 214, 718
\re  Shigeyama T., Nomoto K., Hashimoto M. 1988, A\&A 196, 141
\re Sneden, C., McWilliam, A., Preston, G.W., Cowan, J.J., Burris, D.I., 
Armosky, B.J., 1996, ApJ 467, 819
\re Stuart A., Ord, J.K. 1987, in Kendall's Advanced Theory 
of Statistics, 5th ed. (London: Griffin and Co.)
\re  Tenorio-Tagle G., Bodenheimer P. 1988, ARA\&A 26, 145
\re  Timmes F. X.,  Woosley  S. E., Weaver T. A. 1995, ApJS 98, 617
\re  Tinsley B. M. 1980, Fund. Cosmic Phys. 5, 287
\re Wielen R., Fuchs B., Dettbarn C. 1996, A\&A 314, 438
\re Wilmes M., K$\ddot{\rm o}$ppen J. 1995, A\&A 294, 47

\end{document}